%Paper: 9201060
%From: LINA@lps.umontreal.ca
%Date: Tue, 28 Jan 1992 14:12:50 -0500 (EST)

% This is a TeX file

\magnification=\magstep1
\vsize=22.5truecm
\hsize=16.6truecm
\baselineskip=14pt
\parskip=.1truecm

\font\ad=cmti10
\font\ti=cmbx10 scaled\magstep1
\def \overleftrightarrow #1{\raise 1.5ex \hbox{$\leftrightarrow $}\mkern
-16.5mu #1}

\def\folio{\ifnum\pageno=1\nopagenumbers\else\number\pageno\fi}
\def\sqr#1#2{{\vcenter{\vbox{\hrule height.#2pt
     \hbox{\vrule width.#2pt height#1pt \kern#1pt \vrule
     width.#2pt}\hrule height.#2pt}}}}
\def\square{\mathchoice\sqr63\sqr63\sqr{2.1}3\sqr{1.5}3\,}

\catcode`"=\active
\def\simplequotes{\def\openquote{``\def"{\closequote}}
\def\closequote{''\def"{\openquote}} \def"{\openquote}}

\simplequotes

\line{\hfill UdeM-LPN-TH-59}
\line{\hfill December 1991}

\vskip 2truecm
\centerline{\ti LAX PAIR FORMULATION OF THE W-GRAVITY}
\vskip .2truecm
\centerline{\ti THEORIES IN TWO DIMENSIONS}
\vskip 2truecm
\centerline{Jean-Marc Lina and Prasanta K. Panigrahi}
\vskip.5truecm
\centerline{\ad  Laboratoire de Physique Nucl\'eaire}
\centerline{\ad Universit\'e de Montr\'eal}
\centerline{\ad C.P. 6128, succ``A" }
\centerline{\ad Montr\' eal, Qu\' ebec, Canada}
\centerline{\ad H3C 3J7}
\vskip1truecm
\centerline{\bf Abstract}
The Lax pair formulation of the two dimensional induced
gravity in the light-cone gauge is extended to the more general $w_N$
theories. After presenting the
$w_2$ and $w_3$ gravities, we give a general prescription for an
arbitrary $w_N$ case.
This is further illustrated with the $w_4$ gravity to point out some
peculiarities. The constraints
and the possible presence of the cosmological constants are systematically
exhibited in the zero-curvature condition, which also yields the relevant Ward
identities. The restrictions on the gauge parameters in presence of the
constraints are also pointed out and are contrasted with those of the ordinary
2d-gravity.

\vskip.8in
\centerline{\it Submitted to International Journal of Modern Physics A}

\vfill \eject

\vskip 4cm
\noindent
{\bf I. Introduction}
\vskip .3cm
Induced gravity in two dimensions, arising from the interaction of conformal
matter with the gravitational field, is now being extensively studied, due
mainly to its relevance to string theory$^{[1]}$.
Significant progress has been made in the weak-coupling regime ($c\leq 1$ and
$c>25$, $c$ being the central charge of the matter sector), both in the direct
continuum$^{[2]}$ and the lattice model approaches$^{[3]}$.
Although the non-local effective action, given by
$$
S =\alpha\int
\sqrt{-g}\,\big(\,R\, \,{\square}^{-1}R - \lambda \big)\,d^2x,\eqno(1)
$$
becomes local both in the conformal and the light-cone gauges, the problem of
quantization is much more tractable in the latter case, as was first pointed
out
by Polyakov$^{[1]}$.
In particular, a hidden $sl(2,R)$ Kac-Moody symmetry becomes apparent in this
approach, allowing the extraction of non-perturbative informations about the
theory, e.g the anomalous dimensions of various fields in the presence of
quantum gravitational fields$^{[4]}$.
The light-cone gauge is characterised by the line element
$$ds^2 = dx^+\,dx^- + h(x^+,x^-) {dx^+}^2,\eqno(2)
$$
where the metric $ h(x^+,x^-)$ transforms under the residual coordinate
transformations as
$$\delta h = \epsilon h^\prime + {\dot \epsilon} -h \epsilon^\prime.
\eqno(3)$$
The dot and the prime denote the derivatives with respect to $x^+$
and $x^-$ variables. The simplicity of the light-cone approach can be easily
seen from the equation of constant curvature, which yields a
constraint $$ R=\partial_-^2\,h = -2\lambda, \eqno(4) $$
in contrast to the dynamical Liouville equation in the conformal gauge.
The constraint equation yields the equation of motion for the
induced gravity $\partial_-^3\,h = 0$, allowing the expansion of $h(x^+,x^-)$
as
$$
h(x^+,x^-) = J^-(x^+) -2J^0(x^+) x^- + J^+(x^+) {x^-}^2;\eqno(5)
$$
here $J^-(x^+)$, $J^0(x^+)$ and $J^+(x^+)$ are the currents of the
$sl(2,R)$ Kac-Moody algebra.

\vskip0.6cm
A particularly clear explanation of the geometrical
origin of the current
algebra symmetries in the induced gravity was provided by Polyakov$^{[5]}$, who
demonstrated  how to get diffeomorphisms from the restricted gauge
transformations. Considering a $sl(2,R)$ Lie algebra valued field
$A_\mu\,dx^\mu$, a partial gauge fixing in the $A_-$ sector led to the
transformation of the $A_-^+$ field as the stress-energy tensor of a conformal
field theory. Suitable gauge fixing in the $A_+$ sector then led to the
transformation of the $A_+^-$ component as the light-cone metric; surprisingly,
the gravitational Ward identity emerged as a consistency condition of the gauge
fixing or, equivalently, from the zero-curvature condition. Generalizations to
the $sl(N,R)$ case are now being actively pursued in the literature$^{[6]}$.

In the works of Das et al.$^{[7]}$, the zero-curvature condition, written as a
compatibility equation of a matrix Lax pair, was interpreted as a gauge anomaly
equation. The anomaly equation can, in principle, be integrated to yield a
suitably gauged WZWN action$^{[8]}$. Following this approach, the KdV and the
Boussinesq hierarchies were related with the ordinary and the $w_3$
gravitational
Ward identities respectively. Subsequently, the work of the present authors
incorporated the curvature constraints in this Lax pair formulation of the 2-d
gravity$^{[9]}$.

In the present work, our goal is to analyse the more general
$w_N$ gravities. In particular, we will provide a general
prescription to determine the $A_-$ and the $A_+$ sectors of the $sl(N,R)$ Lie
algebra valued gauge connection such that the compatibility of the Lax pair,
$\partial_- + A_-$ and $\partial_+ + A_+$, will yield the constraints and the
dynamical equations.
We will briefly explain the origin of the constraints in the $w_N$-gravity, and
their relevance for the quantum regime of the theory. The elements of the $A_-$
field, $w_2$ and $w_{k,k>2}$, will transform respectively as the stress energy
tensor and the spin-$k$ quasi-primary fields under the residual gauge
transformations. A conjectured formula, valid to all orders, will be
provided for the infinitesimal variations of these fields under
diffeomorphism. The $A_+$ sector contains variables which transform
as the currents of a conformal field theory. These fields will be interpreted
as
the metrics of the $w_N$ gravity$^{[6,10]}$.

In Sec. II, we give a brief introduction to the $w_N$ algebras and their
connection with the integrable non-linear equations. Our approach is outlined
in
the well studied $sl(2,R)$ and $sl(3,R)$ cases in Sec. III. Sec. IV deals
with the generalisation to an arbitrary $w_N$ gravity. As an illustration, the
$w_4$ gravity is also investigated in this section, pointing out certain
peculiarities in this case and showing explicitly how some non-linear
evolution equations can arise as special cases of the $w_4$ Ward identities.
We
conclude in Sec.V with some final remarks and future directions of work. There
are three appendices; the first one gives the variations of the fields in
$sl(4,R)$, the second lists the variations of the $w_i$'s in $sl(10,R)$ as a
specific example of the conjectured general formula given in Sec. IV. The last
appendix presents the infinitesimal variation of the metric field $h_2$ in some
higher spin $w$-gravity theories.

\vfill\eject

\noindent
{\bf II. $w_N$ algebras and covariant Lax operators} \vskip 0.3cm

A generalization of the Virasoro algebra to the so-called $w_3$
algebra\footnote{$^{1}$}{In the present day terminology, this algebra is known
as the $w_3^{(1)}$ algebra.} was achieved by Zamolodchikov$^{[11]}$. This
algebra
contains a chiral spin-3 conserved current $w_3(z)$ in addition to the
stress-energy tensor $w_2(z)$.
With additional fields, consistent generalizations to higher spin algebras have
been discovered in the literature$^{[12]}$ and these $w_N$ algebras
have appeared in many physical contexts. To quote a few examples, they have
manifested in the gauged WZWN models$^{[13]}$, in the cosets of affine Lie
algebras$^{[14]}$, Toda field theories$^{[15]}$, $2+1$ dimensional Chern-Simons
theories$^{[16]}$ and also in the context of various non-linear integrable
equations$^{[17]}$  e.g. KdV, Boussinesq, etc. It is worth emphasizing that
these
algebras are not Lie algebras because of quadratic defining relations as will
become clear in the course of the text. The $w_N$ algebras have series of
unitary representations characterised by
$$c = (n-1)\bigg(1-{n(n-1)\over
p(p+1)}\bigg),\,\,\,\,p=n+1,n+2,\dots\eqno(6)$$
and hence have been of
interest in the construction of the ``periodic table" of the
conformal-invariant
solutions of the two-dimensional euclidean quantum field theories.
Here, we will be mainly concerned with the classical $w_N$ algebras. In
particular, their connection with the non-linear integrable equations is
pertinent for our work.

\noindent The much studied KdV equation
$${w_2}_t = {w_2}_{xxx} + 6 {w_2} {w_2}_{x},\,\,\,\,\,({\rm
where}\,\,{w_2}_{x}\equiv {\partial {w_2}\over \partial x}\,),\eqno(7)$$
and the related hierarchy of equations can be written as Hamiltonian equations
$${w_2}_t = \{ {w_2}, {\cal H}_n\}_{(l)}\eqno(8)$$
in two distinct ways$^{[18]}$. The two Hamiltonians
$$
{\cal H}_1 = {1\over 2} \int dx\,(2 {w_2}^3 - {{w_2}_x}^2),\,\,\,\,\,\,{\rm
and}\,\,\,\,\,\,\, {\cal H}_2 =  \int dx\, {w_2}^2\eqno(9)$$
give rise to the above equation if the respective Poisson brackets ($l=n$) are
defined as
$$\{ {w_2}(x), {w_2}(y) \}_{(1)} = \partial \delta(x-y),\,\,\,\,\,\,{\rm
and}\,\,\,\,\,\,\,
\{ {w_2}(x), {w_2}(y) \}_{(2)} = ({1\over 2}\partial^3 + {w_2}\partial +
\partial\,{w_2})
\delta(x-y)\eqno(9)$$
where the operator $\partial = {\partial \over \partial x_-}$ acts on all the
objects to its right.  The two Poisson brackets can be recognized as the
abelian current algebra and the Virasoro algebra respectively.

\vskip0.2cm
In Ref.[19], another interesting derivation of the KdV equation, connecting it
to the covariance property of the Hill operator under reparameterisation was
obtained. It has been subsequently extended to other integrable
systems$^{[20]}$. Briefly, the Hill operator
$${\cal L}_{(2)} = \partial^2 + {w_2}(x^-)\eqno(11)$$
transforms covariantly under reparameterisation in the sense that
$$\eqalign{
x^- &\rightarrow f(x^-),\cr
{\cal L}_{(2)} &\rightarrow \big(f^\prime\big)^{-3\over 2}\big(\partial^2 +
{\overline w}_2(x^-) \big)
\big(f^\prime\big)^{-1\over 2}}\eqno(12)$$
where
$${\overline w}_2(x^-) =  \big(f^\prime\big)^{2} {w_2}(f) + {1\over 2}
S_f(x^-),$$
$S_f$ being the Schwarzian derivative, $S_f = {f^{\prime\prime\prime}\over
f^\prime} -{3\over 2} \big({f^{\prime\prime}\over
f^\prime}\big)^2$.
A time (represented by $x^+$) dependent reparameterisation then yields
${w_2}(x^-)\rightarrow {\overline w}_2(x^+,x^-)$ and
$${\dot {\overline w}_2} = {1\over 2} h^{\prime\prime\prime} + 2 {\overline
w}_2
h^{\prime} + {\overline w}_2^\prime h,\eqno(13)$$
where $h$ is defined as ${{\dot f}\over f^\prime}$. The KdV equation in the
standard form is obtained by making the special choice $h = 2 {\overline w}_2$.
It is worth pointing out that the above non-linear evolution equation is
nothing but the gravitational Ward identity for the central charge $c={1\over
2}$ and $h={{\dot f}\over f^\prime}$ is the well known Beltrami equation
defining the light-cone metric.
\vskip 0.2cm

At this point, looking at the infinitesimal variation of ${w_2}(x^-)$ under
$x^-
\rightarrow x^- + \epsilon^-$, the second Poisson bracket mentioned before can
be extracted, assuming ${w_2}(x^-)$ as the generator of diffeomorphism.

\noindent
This procedure, in principle, can be generalized to the higher covariant
operators. Given an $n$th order differential operator of the form $M_n =
\partial^n + \sum_{i=0}^{n-2}u_i\partial^i$, a covariant operator ${\cal
L}_{(n)}$ can be constructed$^{[21]}$ such that it acts on densities of weight
${1-n\over 2}$ and the functions appearing in ${\cal L}_{(n)}$ are the
generators of the $w_N$ algebra. Some examples of ${\cal L}_{(n)}$ are given
below.
$$
\eqalign{
{\cal L}_{(0)} = &1, \cr
{\cal L}_{(1)} = &\partial , \cr
{\cal L}_{(2)} = &\partial^2 + {w_2} ,\cr
{\cal L}_{(3)} = &\partial^3 +4 {w_2} \partial + 2 {w_2}_x + w_3, \cr
{\cal L}_{(4)} = &\partial^4 + 10 {w_2} \partial^2 + 10 {w_2}_x \partial + 9
{w_2}^2 +  3
{w_2}_{xx} +\partial w_3 + w_3\partial + w_4. }\eqno(14)$$
It should be mentioned that the covariant Lax operators ${\cal L}_{(2)}$ and
${\cal L}_{(3)}$ have appeared in the $w_2$ and $w_3$ gravities, in the works
of
Zamolodchikov$^{[22]}$ and Matsuo$^{[23]}$ respectively.

For the sake of completeness, it sould also be pointed out that the above
mentioned Hamiltonian formulations can be related to the symplectic
structures associated with the space of pseudo-differential operators$^{[24]}$.
In particular, the second Poisson bracket structure relevant for the $w_N$
algebras can be extracted following Adler$^{[25]}$.
\vskip 0.2cm

\noindent
Given a Lax operator ${\cal L}_{(n)}$, one defines
$${\partial {\cal L}_{(n)}\over \partial t} = ({\cal L}_{(n)} F)_+ {\cal
L}_{(n)} - {\cal L}_{(n)} ({\cal L}_{(n)} F)_+\eqno(15)$$
where
$$F = \partial^{-1} f_0 + \partial^{-2} f_1 + \dots + \partial^{-n} f_{n-1},$$
and  the $({\cal L}_{(n)} F)_+$ is the
differential part of ${\cal L}_{(n)} F$. Comparing this equation with
$${\partial u_{i} \over \partial t}= \sum_{j=0}^{n-2} D_{ij}^{(2)}
f_j\eqno(16)$$
one finds the Poisson brackets as $\{ u_{i}(x),u_{j}(y)\} =
D_{ij}^{(2)} \delta(x-y)$.
The Lax operators, if written in a covariant form, generate the $w_N$ algebras.

\vskip0.2cm
Although, we will pursue a method involving the matrix valued Lax pairs in
this paper, it turns out that the specific mapping from $M_n$ to ${\cal
L}_{(n)}$ is  of relevance in this formalism. The relationship of the $u_i$'s
and the $w_i$'s appearing in $M_n$ and  ${\cal L}_{(n)}$ respectively is
precisely those of the fields $W_i$'s and the primary fields ${w}_i$'s in this
work. The scalar differential operators will be connected with the matrix Lax
operator $\partial_- +A_-$ to explain this interesting relationship. It is also
worth observing that the inverse mapping appears to determine the $A_+$ sector
and the gauge parameters that generate the gauge preserving transformations.


\vskip 0.6cm
\noindent
{\bf III. $w_2$ and $w_3$ gravities} \vskip 0.3cm
Let us first elaborate on the (matrix) Lax pair formulation of the much studied
ordinary gravity.
\noindent This is done in a manner which can be easily generalized to an
arbitrary $w_N$ algebra. In case of the $w_2$ gravity, one considers the
deformed
$sl(2,R)$ algebra$^{[26]}$
$$[{\bf t}_+,{\bf t}_-] = \lambda_+\lambda_-\,{\bf t}_0,\,\,\,\,[{\bf
t}_0,{\bf t}_\pm] = \pm 2{\bf t}_\pm.\eqno(17) $$
where ${\bf t}_a$'s are represented by the matrices
$$
{\bf t}_-=\left(\matrix{0 & 0\cr \lambda_- & 0\cr}\right),\,\,{\bf
t}_+=\left(\matrix{0 &
\lambda_+\cr 0 & 0\cr}\right),\, \,{\bf t}_0=\left(\matrix{{1} & 0\cr 0 &
{-1}\cr}\right);  \eqno(18)
$$
$\lambda_\pm$ being the so-called spectral parameters. In the limit of
$\lambda_\pm$ going to zero, the algebra reduces to the Poincar\'e algebra in
two dimensions. The Killing metric $g_{ab} = {1\over 2}f_{ac}\,^d \,f_{bd}\,^c$
is degenerate when either $\lambda_+$ or $\lambda_-$ is vanishing. Considering
the reduced Killing metric on $sl(2,R)/U(1)$ given by $2\lambda_+
\lambda_-\,\eta_{ab}$, we define the gauge field $A = A^a\,t_a$
and the line element as the  coset invariant
$$ds^2 = \eta_{ab} A^a A^b.\eqno(19)$$
The gauge fields are the zweibeins and the spin connection.

\noindent The following chiral gauge choice
$$
A_- = \left(\matrix{0 & 0\cr \lambda_- & 0 \cr}\right),\,\,\,\,
A_+ = \left(\matrix{-{1\over 2}h^\prime & \lambda_+\cr \lambda_- h &
{1\over 2}h^\prime\cr}\right), \eqno(20)
$$
yields the constraint equation
$$
R = \partial^2 h = -2\lambda_+ \lambda_-$$
as the integrability condition, $F=dA + A\wedge A =0$.
It can be easily checked that the field $h(x^+,x^-)$ transforms as the
light-cone metric. Furthermore, it is worth pointing out that the same
equations can be obtained as a consistency condition under the gauge
transformations that maintain this gauge choice. Following the remarkable
analysis of Polyakov$^{[5]}$, we can show that the gauge choice
$$
A_- = \left(\matrix{0 & -t\cr
\lambda_- &
 0\cr}\right),\,\,\,\,
A_+ = \left(\matrix{
 -{1\over 2} h^\prime & \lambda_+ - ({1\over 2\lambda_-}+\kappa)
h^{\prime\prime} - t h\cr \lambda_- h &
{1\over 2} h^\prime
}\right)
\eqno(21) $$
yields $\delta t = 2 t
\epsilon^{\prime} + \epsilon t^{\prime} + {1\over
2\lambda_-}\epsilon^{\prime\prime\prime}  $ under the residual gauge
transformations that maintain $A_-$. This variation is easily recognized as the
variation of the stress-energy tensor in a conformal field theory, with a
central charge $c$ equal to ${1\over 2\lambda_-}$. Furthermore, all the
parameters of the gauge variation are not independent:
$$\left\{
\eqalign{  \epsilon^- &= +{1\over \lambda_-}  \epsilon,\cr
\epsilon^0 &= -{1\over 2} \partial
\epsilon,\cr
\epsilon^+ &= -\big({1\over 2\lambda_-} \partial^2  +
t\big)\epsilon;}
\right.\eqno(22)$$
and it can be easily checked that $h(x^+,x^-)$ does transform as the light-cone
metric field:
$$\delta h = -h \epsilon^{\prime} + \epsilon h^{\prime}
 + {\dot \epsilon}.\eqno(23)$$
The consistency requirement for maintaining the
gauge choice or the zero curvature condition $F = 0$ provide the constraint
equation and the gravitational Ward identity given respectively by
$$
\partial^2\,h =
{\lambda_+\over \kappa,}\eqno(24 a)$$
and
$$\partial_+ t =
\bigg((c + \kappa) \partial_-^3 + \partial \,t + t\partial
\bigg) h.\eqno(24 b)$$
The work of Das et al.$^{[7]}$ corresponds to the
case $\kappa = \lambda_+ =0$, for which only the Ward identity exists. It is
worth emphasising that for taking into account the constraint with the possible
presence of a cosmological constant, the parameter $\kappa$ should be
nonvanishing. It should be pointed out that it is precisely the case in the
quantum Ward identity of the 2d-gravity. In this case however, the nonvanishing
$\kappa$ is exactly calculable. In our gauge choice with $\kappa \ne 0$, the
gauge preserving symmetries are restricted to those described by $\epsilon$
such
that $\partial \epsilon =0$ if $\lambda_+ \ne 0$ or $\partial^2 \epsilon =0$ if
$\lambda_+ = 0$. The parameter of the residual coordinate invariance is
restricted as can be oBserved in the work of KPZ$^{[4]}$.

\vskip 0.2cm\noindent
We pursue the analysis with the example of $N=3$ for the pupose of
generalization to an arbitrary $w_N$ case. We choose the following matrix
representation for the $sl(3,R)$ algebra elements(the same as in Ref.[7]):
$$t_0 = \left(\matrix{1 & 0 & 0\cr 0 & -1 & 0\cr 0 & 0 &0\cr}\right)
,\,\,\,\,\,
t_+ = \left(\matrix{0 & 1 & 0\cr 0 & 0 & 0\cr 0 & 0 &0\cr}\right)
,\,\,\,\,\,
t_- = \left(\matrix{0 & 0 & 0\cr 1 & 0 & 0\cr 0 & 0 &0\cr}\right)
,$$
$$t_{00} = \left(\matrix{0 & 0 & 0\cr 0 & 1 & 0\cr 0 & 0 &-1\cr}\right)
,\,\,\,\,\,
t_{++} = \left(\matrix{0 & 0 & 1\cr 0 & 0 & 0\cr 0 & 0 &0\cr}\right)
,\,\,\,\,\,
t_{--} = \left(\matrix{0 & 0 & 0\cr 0 & 0 & 0\cr 1 & 0 &0\cr}\right)
,\eqno(25)$$
$$t_{0+} = \left(\matrix{0 & 0 & 0\cr 0 & 0 & 1\cr 0 & 0 &0\cr}\right)
,\,\,\,\,\,
t_{0-} = \left(\matrix{0 & 0 & 0\cr 0 & 0 & 0\cr 0 & 1 &0\cr}\right).
$$
and define the gauge field $A_-$:
$$A_- = \left(\matrix{0 & -W_2 & -W_3\cr c_1 & 0 & 0\cr 0 & c_2 &0\cr}\right)
\eqno(26)$$
where the $c_i$'s are constants and the $W_i$'s are dynamical fields
with dimension $i$ (the dimension of ${\partial \over \partial x_-}$ is $+1$).
As
we have noticed in Ref.[9], one can set $c_2=1$ without loss of
generality\footnote{$^{2}$}{This constant scales the primary field $w_3$
defined
in the next page; this observation generalizes to the higher spin algebras.}.

\noindent As they stand, the dynamical fields in $A_-$ do not satisfy the
Virasoro symmetry properties mentioned before.
The gauge transformations that preserve Eq.(26) are found
to be described by the functions $\epsilon^-$ and $\epsilon^{--}$ or, more
conveniently, by the arbitrary
functions $\rho_{2}$ and $\rho_{3}$:
$$\left\{
\eqalign{ \epsilon^- &= c_1 \rho_{2} - {c_1\over 2} \rho_{3}^\prime, \cr
\epsilon^{--} &= c_1 \rho_{3},}\right.
\eqno(27)$$
and, defining the $w_i$'s by the following change of
variables
$$
W_2 = w_2, \hskip0.4cm W_3= w_3 -{1\over 2} w_2^\prime,
\eqno(28)$$
$$\left\{
\eqalign{ \epsilon^{0} &= -\rho_{2}^\prime + c_1 \big({1\over 6 c_1}
\partial^2 -{1\over 3} w_2 \big) \rho_{3}, \cr \epsilon^{+} &= -\big({1\over
c_1} \partial^2 + w_2 \big) \rho_{2} + \big({1\over 6 c_1} \partial^3 +{1\over
6} \partial\,w_2 - w_3 \big) \rho_{3}, \cr \epsilon^{00} &= -\rho_{2}^\prime -
c_1 \big({1\over 6 c_1} \partial^2 +{2\over 3} w_2 \big) \rho_{3}, \cr
\epsilon^{0-} &= \rho_{2} + {1\over 2}\rho_{3}^\prime, \cr
\epsilon^{0+} &= -\rho_{2}^{\prime\prime} - c_1 \big({1\over 6 c_1}
\partial^3 -{1\over 6} \partial\,w_2 -{1\over 2} w_2 \partial - w_3 \big)
\rho_{3}.\cr
 }\right.
\eqno(29)$$
Indeed, the gauge preserving transformations lead to the Virasoro and the
spin-3
symmetries in terms of the parameters $\rho_{2}$ and $\rho_{3}$ as defined in
Eq.(27). These symmetries are generated respectively by the energy momentum
tensor $w_2$ and the primary field $w_3$ defined in Eq.(28) since we find that
$\delta_\epsilon(A_-)$ leads to $$\delta \,w_i = \delta(\rho_{2})\,w_i +
\delta(\rho_{3})\,w_i$$ with
$$
\eqalign{ &\delta(\rho_{2})\,w_2  \,\,\,\,\,=\, 2 w_2\, \rho_{2}^\prime\, +
\rho_{2}\,w_2^\prime\,+ {2\over c_1} \rho_{2}^{\prime\prime\prime} ,\cr
&\delta(\rho_{2})\,w_3  \,\,\,\,\, =\, 3 w_3\,  \rho_{2}^\prime\, +
\rho_{2} \,  w_3^\prime,\cr
&\delta(\rho_{3})\,w_2  = 3 w_3\,
\rho_{3}^\prime\, + 2 \rho_{3}\, w_3^\prime,\cr
&\delta(\rho_{3})\,w_3  =  - {1\over12} \big(
2 w_2^{\prime\prime\prime}\,\rho_{3} + 9
w_2^{\prime\prime}\, \rho_{3}^{\prime} + 15  w_2^{\prime}\,
\rho_{3}^{\prime\prime}\, +
10  w_2\, \rho_{3}^{\prime\prime\prime} \big) \cr & \qquad\,\,\,\qquad\qquad-
{2c_1\over 3}( {w_2}^2
\rho_3^{\prime} + w_2  w_2^{\prime}\, \rho_{3})  -
{1\over 6 c_1 }\rho_{3}^{\prime\prime\prime\prime\prime}. }
\eqno(30)$$
We can consistently
construct a Poisson bracket structure between the $w_i$'s, starting with the
Virasoro algebra satisfied by the generator of diffeomorphisms $w_2$:
$$
 \{ w_2(x), w_2(y)\} = \big( c \partial^3 + w_2\partial +
\partial\,w_2 \big) \delta(x-y) \eqno(31)
$$
with $c = {{2 }\over c_1}$. The primary field $w_3$  can be interpreted as the
generator of the spin-3 transformations:
$$
\eqalign{
&\{ w_3(x), w_3(y)\} = -{1\over 12}\big(c \partial^5 + 2 w_2\partial^3
+ 3\partial w_2\partial^2 + 3\partial^2 w_2\partial+ 2\partial^3 w_2  +
{16\over c} {w_2}\partial {w_2} \big)\delta(x-y),   \cr  &\{ w_2(x), w_3(y)\}
=   (w_3\partial + 2 \partial\,w_3) \delta(x-y) \cr &\{ w_3(x), w_2(y)\} =   (2
w_3\partial + \partial\,w_3) \delta(x-y). } \eqno(32)
$$
As can be seen from the Poisson bracket $\{ w_3(x), w_3(y)\}$, there are
non-linear terms which go to zero in the so-called ``classical limit''
$c\rightarrow \infty$. The $A_-$ sector being defined,
$$A_- = \left(\matrix{0 &
-w_2 & -w_3+{1\over2}w_2^\prime\cr c_1 & 0 & 0\cr 0 & 1 &0\cr}\right)
,\eqno(33)
$$
we now study the $A_+$ gauge field. The components $A_+^{a}$ are defined in
terms of the ``metric fields'' $h_i$, $i=2,3$ and some general $n$th order
differential operator, {\sl linear} in the $W_i$'s and denoted by ${\cal
D}_a^{(n)}$: $$
A_+^{a} ={\cal D}_a^{(1 + l(a))}h_2 +
{\cal D}_a^{(2 + l(a))}h_3. \eqno(34)
$$
Here, $l(a)$ represents the number of nonvanishing gauge indices in ``$a$''
e.g.
$l(+)=l(0+)=1$, $l(-)=l(0-)=-1$ ,$l(0)=l(00)=0$, and
$l(++)=-l(--)=2$. Finally, we introduce the possible ``cosmological
constants'' $\lambda_{i}$ through the shift
$$
A_+ \rightarrow A_+ + \lambda_2 {\bf t}_{+} + \lambda_3 {\bf t}_{0+},
\eqno(35)$$
a choice which will be justified shortly. Let us therefore write the gauge
field in the form $$
A_+ = \left(\matrix{ {\dots} & {\lambda_0 +\dots} & {\dots}\cr H_2 & {\dots} &
{\lambda_1 +\dots}\cr H_3 & {\dots} &{\dots}\cr}\right) ,
\eqno(36)$$
where we have emphasized the components $H_i \sim h_i +...$, i.e. the
components $A_+^{-}$ and $A_+^{--}$.  Notice that the gauge fields are, by
construction, {\sl linear} in the fields $W_i$'s and the prescription used to
describe them is easily generalizable to higher rank $sl(N,R)$ algebras. The
significance of the constants is already known$^{[9]}$ for the $sl(2,R)$ and
the
$sl(3,R)$ cases: the $\lambda_i$'s enter in the constraint equations involving
the $h_i$'s, and, for the $w_2$-gravity, $\lambda_2$ can be interpreted as the
cosmological constant.

\noindent The components of $A_+$ are determined in such a way that
the curvature is of the following form $$ \left(\matrix{\Phi_2 & E_1 & E_2 \cr
              0 & \Phi_3 -\Phi_2 & 0 \cr
              0 &      0 &   -\Phi_3
}\right),\eqno(37)$$
with
$$\left\{\eqalign{
\Phi_2 &=\kappa_2\partial^{2} h_2 - \lambda_2,\cr
\Phi_3 &=
\kappa_3\partial^{3} h_3  - \lambda_3}\right.,
\eqno(38)$$
and $E_1 = -\partial_+ w_2 +\cdots,$  $E_2 =
-\partial_+ w_3 +\cdots $. The constants $\kappa_i$ are free
parameters. Together with the ``cosmological constants"
$\lambda_i$, they represent the presence or the absence of the constraints. Let
us notice that the Eqs.(38) are legitimate only if the metric fields $h_i$
have the proper dimensions such that $\partial^{i} h_i$ is dimensionless. In
fact, the zero-curvature condition {\sl alone} does not define a unique $A_+$
since we obtain
$$\left\{ \eqalign{&H_2 = c_1 h_2 + a\,h_3^\prime,\cr
         &H_3 = c_1 h_3,}\right.\eqno(39)$$
where $a$ is free. However, choosing this parameter equal to zero spoils
the symmetry property of $h_2$ under the diffeomorphisms. These
symmetries are easily calculated from Eqs.(26,28) and $\delta A_+ =
\partial_+\epsilon + [A_+,\epsilon]$. Keeping the
interpretation of $h_2$ as a metric field, i.e. Eq.(3), gives $a=-{c_1\over
2}$.
Then, the remaining varations are found to be
$$\eqalign{
\delta(\rho_2)\,h_3  =&\,- 2 h_3\, \rho_2^\prime\, +
\rho_2\,h_3^\prime,\cr
\delta(\rho_3)\,h_2  =& {1\over
6}h_3\,\rho_3^{\prime\prime\prime}-{1\over
4}h_3^\prime\,\rho_3^{\prime\prime}\,  + {1\over
4}h_3^{\prime\prime}\,\rho_3^{\prime}-{1\over
6}h_3^{\prime\prime\prime}\,\rho_3+{4 \over 3c}w_2\,h_3\,\rho_3^\prime -
{4\over 3c}w_2\,h_3^\prime\,\rho_3\cr
\delta(\rho_3)\,h_3  =& -h_2\,\rho_3^{\prime}+ 2 h_2^{\prime} \rho_3 +
{\dot\rho_3}. }
\eqno(40)$$
Finally, the dynamical equations $E_{1,2}=0$, obtained from the zero-curvature
condition are the following:
%%%%%%%%%%%%%%%%%%%%%%%%%%%%%%%%%%%%%%%%%%%%%%%%%%%%%%%%%%%%%%%%%%%%%%%%%%%
$$
\partial_+ w_2 = \big( (c + \alpha_2)\partial^3  +
w_2\partial + \partial\,w_2\big) h_2 + \big(\beta_2 \partial^4
+ (w_3\partial  +
2\partial\, w_3 \big) h_3,
\eqno(41a)$$
$$
\eqalign{
\partial_+ w_3 = &\big(\alpha_3 \partial^3 +
2 w_3\partial + \partial\, w_3 \big) h_2  \cr - &{1\over
12}\big((c+\beta_3)\partial^5 + (2 \partial^3\, w_2
+3 \partial^2\, w_2 \partial
+3 \partial\, w_2 \partial^2
+2 w_2 \partial^3) + {16\over c} {w_2}\partial\,{w_2} \big) h_3.
}\eqno(41b)$$
%%%%%%%%%%%%%%%%%%%%%%%%%%%%%%%%%%%%%%%%%%%%%%%%%%%%%%%%%%%%%%%%%%%
The $\alpha_i$'s and $\beta_i$'s are linear
combinations\footnote{$^{3}$}{Notice that the explicit expressions for the
$\alpha_i$'s and the $\beta_i$'s are irrelevant in this classical analysis.} of
the $\kappa_i$'s. It is remarkable that they only occur in the shift of the
central charges, as in the $sl(2,R)$ case. Let us recall that this system
leads to the Boussinesq equation by setting $h_2=0$ and $h_3={1\over 12}$. The
particular case $\kappa_i=\lambda_i=0$ reproduces the analysis of Das et
al{$^{[7]}$}. \vskip 0.1cm

The previously mentioned
covariant operators ${\cal L}_{(2)}$ and ${\cal L}_{(3)}$ can be easily seen to
correspond to the matrix determinant of the gauge covariant operators  ${\cal
D}_- = \partial + A_-$ respectively for the gauge choice of Eq.(21) and
Eq.(33). This observation will be very useful in the next section where we
investigate the general $sl(N,R)$ case.
We will
adopt the approach of partial gauge fixing {\sl \`a la} Polyakov in obtaining
the $w_N$ algebras in the $A_-$ sector and the corresponding metric fields in
the $A_+$ sector.
The guiding principle in the $A_-$ sector is to keep the correspondance between
the scalar Lax operator ${\cal L}_{(N)}$ and the matrix covariant operator
${\cal
D}_-$. Although this approach seems lacking in physical motivations, recent
works
have illustrated connection of this method with 2+1 dimensional Chern-Simons
theories$^{[16]}$. Hence, assuming validity of the Chern-Simons approach for
the general $sl(N,R)$ case, we will proceed with the present analysis.


\vskip 0.6cm
\noindent
{\bf IV. General Lax pair: from $sl(3,R)$ to $sl(N,R)$}
\vskip 0.3cm

This section presents a general algorithm to calculate the matrix Lax
pair associated with $sl(N,R)$. This is done in three steps. First, the $A_-$
sector of the gauge field is fixed in such a way that the remaining degrees of
freedom in $A_-$, the so-called $w_i$'s, are Virasoro quasi-primary fields
under
the remaining gauge preserving transformations. The requirement of
quasi-primary nature of the $w_i$'s avoids
the complicated non-linearities that would have occurred in fixing the
interaction terms in $A_+$ with strictly primary $w_i$'s.
In the second step, we define the $A_+$ sector by introducing a new family of
fields, the $h_i$'s, which are analogs of the metric field for the $w_N$
generators. Our objective now is to obtain, as in the $sl(2,R)$ case, the
relevant Ward identities and the consistent constraints involving the $h_i$'s
from the zero-curvature condition. This procedure determines most of the
components of $A_+$; the remaining freedom is related with the constraint
equations themselves (analogs of the $\kappa$ and $\lambda$ of the $sl(2,R)$
case) and the symmetry properties of the $h_i$'s under diffeomorphisms. This
leads to the last step, similar to the first one, where we require the fields
$h_i,i>2$ to transform like currents under the fundamental Virasoro symmetry:
$$ \delta(\rho) h_{i,i>2} = -(i-1)h_i\,\rho^\prime +
h_i^\prime \rho.  \eqno(42)$$
For $h_2$, we want to maintain Eq.(3), at least for the finite subgroup of
diffeomorphism. This requirement and Eq.(42) imply that $\partial^i
h_i$ are dimensionless quantities and, hence, the constraint equations
$$\Phi_i \equiv \kappa_i \partial^i h_i - \lambda_i \sim 0,\,\,\,\,\,\forall i
=2,3,...,N
\eqno(43)$$
are meaningful. This gives us the opportunity to introduce $N$ ``cosmological
constants'' $\lambda_i$ and $N$ parameters $\kappa_i$. Vanishing of the
$\kappa_i$'s and the $\lambda_i$'s eliminates these constraints.

\vskip0.2cm
We now propose an algorithm that realizes for any $sl(N,R)$ algebra the gauge
choice that satisfies the previous conditions. Keeping the form
of the gauge suggested by the $sl(2,R)$ and $sl(3,R)$ cases, let us write
$$ A_-
= \left(\matrix{    0   & -W_2 & -W_3    &  -W_4     &\dots &{} &-W_N\cr
                    c_1 & 0    &  \dots  &   \dots   &{}    &   & 0 \cr
                    0   & 1    &  0      &   \dots   & {}   &{} & 0 \cr
                    0   & 0    &  1      &  0        &\dots &{} & 0 \cr
               \vdots   & {}   &  {}     &  {}       & {}   &{} & 0 \cr
                    0   & 0    &  {}     &  {}       &{}    & 1 & 0 \cr
}\right),\eqno(44)$$ $$
A_+ =\left(\matrix{\dots & \dots &  \dots&   \dots    &  \dots &{}\cr
                    H_2   & \dots  &   \dots  &    \dots  &   {}  & {} &   {}
\cr
                    H_3   & \dots  &   \dots  &    {}  &   {}  & {} &   {}  \cr
                    H_4   &  \dots &   {}  &  {}  &   {}  &   {} &   {}  \cr
                    \vdots   & {}  &  {}  &  {}  &   {}  &   {} &   {}  \cr
                    H_N   & \dots  &  {}  &  {}  &   {}  &   {} &   {}  \cr
}\right)
. \eqno(45)$$
We will put $c_1 = 1$ for convenience; this parameter scales the central
charge.
The other constant components of $A_-$ could have been chosen different from
unity; however, rescaling of the primary fields in $A_-$ eliminates these
parameters in both the resulting Poisson brackets and the dynamical equations.

\noindent The $N-1$ fields $W_i$ and $H_i$ have been considered in the
Refs.[6,7]. It is well-known that the $W_i$'s are {\sl not} quasi-primary
fields and the $H_i$'s do not transform as definite spin currents. Let us
consider the following change of variables:
$$\left\{
\eqalign{
W_i &= \sum_{j=2}^{i}(-1)^{i-j}\, A_{ij}(N)
\,\partial^{i-j} w_j, \,\,\,{\rm with}\,\,\, A_{ij}(N) ={{i-1\choose
i-j}{N-j\choose i-j}\over {i+j-1\choose i-j}}\cr
H_i &=
\sum_{j=i}^{N} B_{ji}(N) \,\partial^{i-j} h_j, \,\,\,{\rm with}\,\,\, B_{ij}(N)
= (-1)^{i-j}{{i-1\choose i-j}{N-j\choose i-j}\over {2i-2\choose i-j}},\cr}
\right.\eqno(46)$$
that we will justify later. The gauge field $A_-$
being thus completely defined, the zero-curvature condition yields the
following
recurrent relations on the matrix elements $a_{ij}$ of $A_+$
$$\left\{
\eqalign{ &a_{i,j+1} - a_{i-1,j} =
a_{i,j}^\prime - H_iW_j,\cr &a_{k,k} =  \sum_{i=2}^{k}( a_{i,i-1}^\prime -
H_iW_{i-1}) -\sum_{i=2}^{k}{N+1-i\over N}( a_{i,i-1}^\prime - H_iW_{i-1}),\cr
&a_{N-i,N} = - (a_{N-i+1,N}^\prime - H_{N-i+1}W_N).
 }
\right.\eqno(47)$$
These equations determine all the elements of $A_+$ in terms of the
$N-1$ fields $H_i$'s. Let
us ignore the constraints for the moment and concentrate on the
symmetries. It is rather obvious that the issue of the gauge preserving
symmetries $\delta A_- \equiv \partial_- \epsilon + [A_-,\epsilon]$ is similar
to the zero-curvature condition since, on the constant elements of the gauge
field in Eq.(44), the gauge preserving condition amounts to solving $\partial_-
\epsilon - \partial_+ A_- + [A_-,\epsilon]=0$. Therefore, the matrix $\epsilon$
is determined by its first column
$$
\epsilon =\left(\matrix{\dots & \dots &  \dots&   \dots    &  \dots &{}\cr
                    \epsilon_2   & \dots  &   \dots  &    \dots  &   {}  & {} &
  {}  \cr
                    \epsilon_3   & \dots  &   \dots  &    {}  &   {}  & {} &
{}  \cr
                    \epsilon_4   &  \dots &   {}  &  {}  &   {}  &   {} &   {}
\cr
                    \vdots   & {}  &  {}  &  {}  &   {}  &   {} &   {}  \cr
                    \epsilon_N   & \dots  &  {}  &  {}  &   {}  &   {} &   {}
\cr }\right)
, \eqno(48)$$
and the recurrent equations (47) can be used to compute all the other
components. The knowledge of the $\epsilon$ leads to the proper variations of
the
$w_i$'s and of the $h_i$'s.
It is remarkable that the generalization of Eq.(27), in order to fulfill the
symmetry requirements previously described, is the following redefinition of
the gauge parameter:
$$
\epsilon_i =
\sum_{j=i}^{N} B_{ji}(N) \, \partial^{i-j} \rho_j, ,\eqno(49)$$
with the matrix $B(N)$ defined in Eq.(46). More precisely, looking only at the
variations induced by $\rho_2$, the gauge transformations lead to the
following expressions $$\left\{
\eqalign{
&\delta w_2 = 2\,w_2\,\rho_2^\prime + w_2^\prime \rho_2 +
C(N)\,\rho_2^{\prime\prime\prime},\cr &\delta w_{2k+1} =
{(2k+1)}\,w_{2k+1}\,\rho_2^\prime + w_{2k+1}^\prime \rho_2 + \sum_{l=1}^{k-1}
C_{2k+1}^l(N)\,\Omega_l^{2k+1},\cr &\delta w_{2k} =
{2k}\,w_{2k}\,\rho_2^\prime + w_{2k}^\prime \rho_2 + \sum_{l=1}^{k-1}
C_{2k}^l(N)\,\Omega_l^{2k}.\cr} \right.\eqno(50)$$
We have defined the central charge
$C(N)={(N-1)N(N+1)\over12}$ and
$$\Omega_l^{q} = \sum_{i=0}^{2(l-1)}
{(-1)^i\sigma_{i}(l)\over {2q-4l+i-1\choose i}} w_{q-2l}^{(i)}\,
\rho_2^{(2l-i+1)}, \eqno(51)$$
where $\sigma_{i}(l)$ are given by:$$
\sigma_{0}(l)=1,\,\,\sigma_{1}(l)=
2(l-1)(2l+1),\,\,\sigma_{2}(l)=l(l-1)(2l+1)(2l-3),
$$
and,
$$\sigma_{i\ge 3}(l)={{2l+1\choose i}{2(l-1)\choose i-3}{2l+1-i\choose 3}\over
{i\choose 3}}.\eqno(52)$$
Let us note that the
derivatives appearing on the parameter $\rho_2$ are always
greater or equal to three which implies that we have defined the gauge field
$A_-$ in terms of quasi-primary fields. Futhermore, the dependance on $N$, i.e.
on the central charge, is completely contained in the coefficients
$C_{k}^l(N)$.
A lengthy calculation leads to the following conjecture$^{[28]}$
$$ C_{k}^l(N) =  {(l-1)! \over2^l
(2l+1)}{N-k+2l\choose 2l}{k-l-2\choose l-1} {\big(2(k-l)-1\big)l\,N +
(k-l)(k-l-1) + l^2 \over \prod_{i=1}^{l}\big(2(k-i)-1\big)} \eqno(53)$$
Finally, let us mention that we obtain the correct symmetry varitions of the
$h_i$'s for $i>2$:  $$ \delta (\rho_2) h_{i>2} = -(i-1)h_i\,\rho_2^\prime +
h_i^\prime \rho_2.$$
However, the variation of $h_2$ involves higher derivative
terms which vanish only in the finite subgroup of the diffeomorphisms. This
point will be illustrated below with some particular cases $N\ge 4$.

\vskip 0.2cm
The ``cosmological constants'' can now be added into the $A_+$ sector. We
consider the following shift of the previous gauge field $A_+$:
$$
A_+\rightarrow A_+ +  \left(\matrix{
0
& \lambda_2-\kappa_2\partial^2h_2
&0
&0
&  \dots
&0\cr
0   & 0
&   \lambda_3-\kappa_3\partial^3h_3
&    0 &   \dots
&0  \cr
0   & 0  &   0  &    \lambda_4-\kappa_4\partial^4h_4  &   \dots  &
{}  \cr
\vdots    &  {}  &   {}  &  {}  &   {}  &  \vdots  \cr
0    &  \dots  &   {}  &  {}  &   0  &   \lambda_N-\kappa_N\partial^Nh_N  \cr
0   & \dots  &  {}  &  {}  &   {}  &   0   \cr
}\right).
\eqno(54)$$
The resulting zero curvature equations exhibit the expected dynamical
equations on the fields $w_k$ with, as already noticed in the $N=2$ and $N=3$
cases, a shift in the coefficients of the higher derivative of the $h_i$'s. The
diagonal entries of $F_{+-}$ now exhibit the expected $N-1$ equations $\Phi_i
\sim 0$.

\noindent It is easy to show that the presence of
constraints restricts the gauge preserving symmetry to that of the finite
subgroup of the diffeomorphisms. Of course, such restriction is also seen
in the quantum regime where the relevance of the constraints is well understood
in terms of the ghost system. Let us only mention that, similar to Eq.(5), the
operators $h_i$'s are expanded in terms of $sl(N,R)$ current operators
$$h_i =
\sum_{n=2}^{2 i -2} {x^-}^n\,J_i^{(n)}(x^+). \eqno(55)$$
Such expansions are
legitimate by the equations $\partial^{2i-1} h_i =0$ that generalize the
constant curvature equation and are easily obtainable from the dynamical
equations with the $w_i$'s equal to zero. Then, the equations $\Phi_i  \sim 0$
lead to the following system of constraints $$J_i^{(i)} \sim {\rm const,}
\qquad
{\rm and}\qquad\left\{ \eqalign{ &J_i^{(i+1)} \sim 0,\cr
&J_i^{(i+2)} \sim 0,\cr
&\,\,\vdots\cr
&J_{i,i>2}^{(j)} \sim 0,\cr
&\,\,\vdots\cr
&J_i^{(2i-2)} \sim 0.
 }, i=2,3,...,N
\right.\eqno(56)$$
Dimensional analysis of this set of constraints completely describes the
weight spectrum of the necessary ghost systems. The resulting central charge
contribution is$^{[10,28]}$ $c_{ghost} = -n^2(n+1)^2 + n(n+1) +2$.

\vskip 0.3cm
\noindent Let us illustrate the previous results with the case of $N=4$. The
definitions in Eq.(46) give $$\eqalign{ &W_2 = w_2, \cr
&W_3 = w_3 -w_2^\prime, \cr
&W_4 = w_4 -{1\over 2} w_3^\prime +{3\over 10} w_2^{\prime\prime};
}\eqno(57)$$
and
$$\eqalign{
&H_2 = h_2-h_3^\prime +{1\over 5}h_4^{\prime\prime}, \cr
&W_3 = h_3 -h_4^\prime, \cr
&W_4 = h_4.
}\eqno(58)$$
The gauge is then uniquely determined through the relations (47) and, finally,
the shift (54) introduces the cosmological constants.
Inspecting the gauge transformations that maintain the gauge so defined, we
obtain the  $c={5}$ Virasoro algebra for $w_2$ together with the
following infinitesimal variations:
$$\eqalign{
&\delta(\rho_2)\,w_3  =\,3 w_3\, \rho_2^\prime\, +
\rho_2\,w_3^\prime,\cr
&\delta(\rho_2)\,w_4  =\,4 w_4\, \rho_2^\prime\, +
\rho_2\,w_4^\prime + {9\over 10} w_2\, \rho_2^{\prime\prime\prime},\cr
&\delta(\rho_2)\,h_2  =\,- h_2\, \rho_2^\prime\, +
\rho_2\,h_2^\prime - {9\over 10} h_4\, \rho_2^{\prime\prime\prime}+ {\dot
\rho_2},\cr &\delta(\rho_2)\,h_3  =\,- 2 h_3\, \rho_2^\prime\, +
\rho_2\,h_3^\prime,\cr
&\delta(\rho_2)\,h_4  =\,- 3 h_4\, \rho_2^\prime\, +
\rho_2\,h_4^\prime,}\eqno(59)$$
and the higher spin variations listed in the Appendix A. The variation
$\delta(\rho_2)\,w_4$ is a consequence of our
gauge choice linear in the $w_i$'s. However, we can redefine the
parameterisation of the gauge and introduce a non-linear term in  $W_4$:
$$W_4  =
w_4 -{1\over 2} w_3^\prime +{3\over 10} w_2^{\prime\prime} +{9\over
100} {w_2}^2\eqno(60)$$
such that $w_4$ is now primary, like $w_3$. This remark leads us to the
justification of the parameterisation of $W_i$ in Eq.(46). It mainly relies on
a
theorem due to Itzykson et al.$^{[21]}$ in the context of the scalar Lax
operator ${\cal L}_{(N)}$; the matrix determinant being the operation that
connects our approach to theirs. Similar to Ref[21], our gauge is thus
parameterized with quasi-primary fields. The previous redefinition (60)
illustrates the missing part of the theorem, i.e. the non-linear contribution
in
$W_i$ that makes {\sl all} the $w_{i>2}$ primary. Using the variations
in Eq.(50), the expression of $W_4$ given by Eq.(60) can be generalised to
higher $N\ge 4$:  $$W_4  =
w_4 + \sum_{j=2}^{3}(-1)^{4-j}\, A_{4j}(N)
\,\partial^{4-j} w_j \,+ {C_4^1(N)\over
C(N)} w_2^2,\eqno(61)$$
Further generalizations to higher dimensional $W_i$'s are certainly possible.
 For example,
$$\eqalign{
&W_5  =
w_5 + \sum_{j=2}^{4}(-1)^{5-j}\, A_{5j}(N)
\,\partial^{5-j} w_j \,+ {C_5^1(N)\over
C(N)} w_2 w_3,\cr
&W_6  =
w_6 + \sum_{j=2}^{5}(-1)^{6-j}\, A_{6j}(N)
\,\partial^{6-j} w_j \,+ {C_6^1(N)\over
C(N)} w_2 w_4 + {C_6^2(N)\over
4\, C(N)}\big(4 w_2 w_2^{\prime\prime} - 5 {w_2^\prime}^2 \big)
\cr & \quad\qquad + {C_6^1(N)\,C_4^1(N) - 9 C_6^2(N)\over 6 {C(N)}^2} {w_2}^3
,\cr
&W_7  =
w_7 + \sum_{j=2}^{6}(-1)^{7-j}\, A_{7j}(N)
\,\partial^{7-j} w_j \,+ {C_7^1(N)\over
C(N)} w_2 w_5 + {C_7^2(N)\over
21\, C(N)}\big(21 w_3 w_2^{\prime\prime} - 35 w_3^\prime{w_2^\prime} \cr
& \quad\qquad+ 10 w_3^{\prime\prime} w_2 \big)   +
{7C_7^1(N)\,C_5^1(N) - 24 C_7^2(N)\over 14\,{C(N)}^2} {w_2}^2 w_3,\cr & {\rm
etc...} }\eqno(62)
$$
A complete formula for this parameterisation of the $W_k$'s in terms of $w_2$
and
{\sl primary} $w_i$'s is under investigation.

\noindent The parameterisation of $H_i$ in
Eq.(46) is more intriguing since, as shown in Ref.[21], the matrices
$A_{ij}(N)$
and $B_{ij}(N)$ satisfy $A.B=1$.  Whereas the fields $h_i$'s satisfy the
symmetry
conditions described previously, the Virasoro variation of the metric field
$h_2$
exhibits a contribution that only vanishes on the finite subgroup of the
diffeomorphisms. This observation is also true for higher $w_N$ gravities (see
Appendix C). However, it is possible to introduce coupling terms in the field
$H_2$
$$H_2 = h_2-h_3^\prime +{1\over 5} \big( \partial^2 + {9\over
2}w_2\big)h_4\eqno(63)$$
such that the field $h_2$ now exactly transforms as in Eq.(3).

\vskip0.2cm \noindent Let us come back to our $N=4$ example and consider the
zero curvature equations. They exhibit the constraints on the $h_i$'s and the
following dynamical equations
%%%%%%%%%%%%%%%%%%%%%%%%%%%%%%%%%%%%%%%%%%%%%%%%%%%%%%%%%%%%%%%%%%%%%%%%%%%  $$
$$
\eqalign{
\partial_+ w_2 = &\bigg( (5+\alpha_2)\partial^3  +
w_2\partial + \partial\,w_2\bigg) h_2 +
\bigg(\beta_2 \partial^4 + (w_3\partial  +
2\partial\, w_3) \bigg) h_3 \cr &+
\bigg(\gamma_2\partial^5 + {9\over 10} \partial^3 \,w_2 + (w_4 \partial
+3  \partial\, w_4)  \bigg) h_4 }\eqno(64a)$$
%%%%%%%%%%%%%%%%%%%%%%%%%%%%%%%%%%%%%%%%%%%%%%%%%%%%%%%%%%%%%%%%%%%%%%%%%%%
$$\eqalign{ \partial_+ w_3 = &\big(\alpha_3\partial^4 +  (2 w_3\partial +
\partial\, w_3) \big) h_2
-{1\over 5}\bigg((5+\beta_3)\partial^5 + (2
\partial^3\, w_2  +3 \partial^2\, w_2 \partial
+3 \partial\, w_2 \partial^2 \cr &
+2 w_2 \partial^3)
- 10 ( w_4 \partial + \partial\, w_4) + 5{w_2}\partial\,{w_2}
\bigg) h_3  - {1\over 10}\,
\bigg(\gamma_3\partial^6
+ 5 \partial^3 \, w_3
+5  \partial^2 \, w_3 \partial
 \cr & +3 w_3 \partial^3) + 5\, w_2 \partial\, w_3 \bigg) h_4
}\eqno(64b)$$
and
%%%%%%%%%%%%%%%%%%%%%%%%%%%%%%%%%%%%%%%%%%%%%%%%%%%%%%%%%%%%%%%%%%%
$$\eqalign{
&\partial_+ w_4 = \bigg(\alpha_4\partial^5 +
 {9 \over 10} w_2 \partial^3
+ 3 w_4 \partial
+ \partial \,w_4 \bigg) h_2 -{1 \over 10}\bigg(
\beta_4\partial^6 + \partial^3\, w_3
+3\partial^2\,w_3 \partial
+5\partial\,w_3 \partial^2 \cr &\,\,\,
+5 w_3 \partial^3
+5 w_3  \partial\,w_2 \bigg) h_3  \,+ {1\over 100}
\bigg((5+\gamma_4)\partial^7
+ 3\partial^5\,w_2 +5 \partial^4\,w_2\partial
+6\partial^3\,w_2\partial^2 \cr &\,\,\, + 6\partial^2\,w_2\partial^3
+5 \partial\,w_2\partial^4 + 3 w_2\partial^5  + 10 \big( \partial^3\,w_4 +
2\partial^2\,w_4\partial +2 \partial\,w_4\partial^2 + w_4\partial^3 \big) \cr
&\,\,\, +
 5 \big( \partial^3\,{w_2}^2 + {w_2}^2\partial^3 +
4\partial\,{w_2}\partial\,{w_2}\partial + 4 {w_2}\partial^3{w_2}\big) - 75 w_3
\partial \,w_3 + 100 \big(w_2 w_4 \partial + \partial\,w_2 w_4\big) \bigg)
h_4 }\eqno(64c)$$
We notice that equations (64a,b) with
$h_2=h_3=0$, $h_4={1\over 2}$ and $w_4=w_2^2$ lead to a coupled KdV system:
$$\left\{ \eqalign{
&\partial_+ w_2-{1\over 2}w_2^{\prime\prime\prime}
-3 w_2^\prime w_2 =0,\cr
&\partial_+ w_3-{1\over 4}w_3^{\prime\prime\prime}
-{1\over 4} w_3^\prime w_2 =0,
}  \right. \eqno(65)$$
whereas Eq.(64c) becomes the following constraint equation:
$$
{1\over 100}w_2^{\prime\prime\prime\prime\prime}+{1\over
20}{w_2^2}^{\prime\prime\prime}-{3\over 5}w_2w_2^{\prime\prime\prime}
+{1\over 3}{w_2^3}^{\prime} = {1\over 8} {w_3^2}^{\prime}.
\eqno(66)$$

\vfill\eject

\noindent
{\bf V. Conclusion} \vskip 0.3cm

In this note, we studied the Lax pair formulation of the $w_N$ gravity
theories; the analysis being strictly at the classical level. One of the
matrix Lax operator $\partial + A_-$ yielded the basic differential operator
associated with the $n$th reduction of the KP hierarchy, after taking the
matrix determinant or, equivalently, solving the differential equation
$(\partial + A_-)\Psi = 0$ where $\Psi$ takes its value in a jet
bundle$^{[6]}$. Hence, interestingly, the partial gauge fixing of the $A_-$
sector, designed for obtaining fields transforming as higher spin objects under
the residual gauge transformations, turned out to be connected with the process
of covariantization of this $n$th order differential operator. It will be of
interest to find out the geometry behind the other members of the $w^{(l)}_N$
algebras. Although these fields were chosen to be quasi-primary for the sake of
convenience, the generalisation to primary fields can be achieved by using the
above mentioned connection with the scalar covariant operators. We
gave a conjectured formula, valid to all orders, for the
infinitesimal variations of the $w_i$'s under diffeomorphism. By
studying these variations, it is possible, in principle, to
construct the primary fields out of the quasi primary ones. It is of
deep interest to obtain an all order formulation for the gauge defined by
Eq.(44) in terms of primary fields; this problem is currently under
investigation. Furthermore, the work of Zamolodchikov has shown the connection
of
these covariant operators (in the absence of higher spin fields) with the
induced Liouville action, which indicates a large-N analysis of this problem
can
be carried out using the matrix valued connections thereby clarifying the
relationship between the continuum and the lattice model approaches to the two
dimensional gravity. The $A_+$ sector, related with the $sl(N,R)$ current
algebra, has been determined modulo scalings of the metric fields $h_i$'s by
constant parameters, by requiring $h_{i>2}$ to transform as currents under the
Virasoro symmetry and $h_2$ to transform as the light-cone metric. The scaling
freedom is tied with the level of the Kac-Moody algebra. The relationship of
the
original $H_i$'s and the $h_i$'s is of great interest: it happens to be the
inverse of the transformation relating the noncovariant fields in $M_{(N)}$ to
the covariant ones in ${\cal L}_{(N)}$. The same connection also holds for the
gauge parameters; namely the infinitesimal parameters associated with the
Virasoro and the higher spin symmetries and the $\epsilon^-$'s appearing in the
gauge transformations are connected to each other by an identical relation. We
suspect that this intriguing connection originates from the coadjoint orbits of
the Lax operator. This point needs further study for clarification.

\noindent
The other entries in $A_+$ were determined by demanding that the zero curvature
condition should yield the relevant constraints and the Ward identities. It is
worth mentionning that the same equations follow from the requirement of the
consistency of the gauge fixing. This is natural because the Lie derivative of
a gauge field can be described as a gauge transformation if the curvature
tensor vanishes. The $F_{+-}=0$ condition immediately suggests a connection
with the Chern-Simons theories, where similar condition appears as the Gauss
law
constraint. In fact, this connection has been already illustrated for the $w_2$
and $w_3^{(2)}$ cases$^{[16]}$; the proof for the $w_N$ case is of obvious
interest. In light of the beautiful geometry associated with the compact
non-abelian Chern-Simons theories, this point needs deeper understanding.
Another open question is the construction of the covariant action like that of
the KPZ analysis of the $sl(2,R)$ case and understanding of the precise
relationship between the non-linear hierarchies appearing in the continuum
approach and those of the matrix models.

\vskip 1cm
\centerline {Acknowledgements}
\bigskip

This work was supported in part by funds provided by the Natural
Sciences and Engineering Research Council (NSERC) of Canada.
\bigskip
%%%%%%%%%%%%%%%%%%%%%%%%%%%%%%%%%%%%%%%%%%%%%%%%%%%%%%%%%%%%%%%%%%%%%%%%%%%%
\bigskip

\noindent
{\bf Appendix A}:

\noindent
The infinitesimal higher spin variations of the $w_i$'s and the $h_i$'s
in the $sl(4,R)$ case are the following:

$$\eqalign{
\delta(\rho_3)\,w_2  =&
3 w_3 \rho_3^\prime + 2 \rho_3 w_3^\prime,\cr
\delta(\rho_3)\,w_3  =&
4 w_4 \rho_3^\prime +
2  w_4^\prime\rho_3 + 2  w_2\,\rho_3^{\prime\prime\prime} +
3 w_2^\prime\,\rho_3^{\prime\prime} + {9\over 5} w_2^{\prime\prime}
\rho_3^{\prime} + {2\over 5} w_2^{\prime\prime\prime}
\rho_3 \cr &- ( {w_2}^2 \rho_3^\prime + w_2  w_2^\prime \rho_3) -
\rho_3^{\prime\prime\prime\prime\prime}, \cr
\delta(\rho_3)\,w_4  =&
-{7\over 5} w_3 \rho_3^{\prime\prime\prime} - {7\over 5} w_3^\prime
\rho_3^{\prime\prime} - {3\over 5} w_3^{\prime\prime} \rho_3^{\prime} -
{1\over 10} w_3^{\prime\prime\prime} \rho_3 \cr & - {1\over 2}(w_2 w_3
\rho_3^\prime + w_3^\prime w_2 \rho_3),\cr
\delta(\rho_3)\,h_2  =&
{2\over 5} h_3 \rho_3^{\prime\prime\prime}
-{3\over 5} h_3^\prime \rho_3^{\prime\prime} +
{3\over 5} h_3^{\prime\prime} \rho_3^{\prime} -
{2\over 5} h_3^{\prime\prime\prime} \rho_3 \cr &-
 {1\over 2}(h_4^\prime w_3 \rho_3 + h_4 w_3^\prime \rho_3) +
h_3 w_2 \rho_3^\prime - h_3^\prime w_2 \rho_3,\cr
\delta(\rho_3)\,h_3  =&
2 h_2^\prime \rho_3 - h_2 \rho_3^\prime + {\dot \rho_3} \cr &
+{1\over 2} h_4 \rho_3^{\prime\prime\prime}  -{1\over 2} h_4^\prime
\rho_3^{\prime\prime} + {3\over 10} h_4^{\prime\prime} \rho_3^{\prime} -
{1\over 10} h_4^{\prime\prime\prime} \rho_3 \cr & +
 {1\over 2}(h_4^\prime w_2 \rho_3 + h_4 w_2^\prime \rho_3),\cr
\delta(\rho_3)\,h_4  =& 2 h_3^\prime \rho_3 - 2 h_3 \rho_3^\prime
,\cr
}$$
\vfill\eject
$$\eqalign{
\delta(\rho_4)\,w_2  =&
4 w_4 \rho_4^\prime + 3 \rho_4 w_4^\prime +{9\over 10}( w_2
\rho_4^{\prime\prime\prime} + 3 w_2^\prime
\rho_4^{\prime\prime} + 3 w_2^{\prime\prime}
\rho_4^{\prime} + 3 w_2^{\prime\prime\prime}
\rho_4) ,\cr
\delta(\rho_4)\,w_3  =&
-{1\over 10} (14 w_3 \rho_4^{\prime\prime\prime} + 28 w_3^\prime
\rho_4^{\prime\prime} + 20 w_3^{\prime\prime} \rho_4^{\prime}+ 5
w_3^{\prime\prime\prime} \rho_4)-{1\over 2}( w_2 w_3\rho_4 ^\prime
+ w_2  w_3^\prime \rho_4),\cr
\delta(\rho_4)\,w_4  =&  {1\over 10} (w_4 \rho_4^{\prime\prime\prime}
+9 w_4^\prime \rho_4^{\prime\prime} + 5 w_4^{\prime\prime}
\rho_4^{\prime} +6 w_4^{\prime\prime\prime} \rho_4 )\cr & + {7\over
25}w_2  \rho_4^{\prime\prime\prime\prime\prime}+ {7\over
10}w_2^\prime  \rho_4^{\prime\prime\prime\prime} + {21\over
25}w_2^{\prime\prime}  \rho_4^{\prime\prime\prime} + {14\over
25}w_2^{\prime\prime\prime}  \rho_4^{\prime\prime} + {1\over
5}w_2^{\prime\prime\prime\prime}  \rho_4^{\prime} + {3\over
100}w_2^{\prime\prime\prime\prime\prime}  \rho_4 \cr
& -{3\over 4} ( w_3^2 \rho_4^{\prime} + w_3 w_3^\prime \rho_4) + 2 w_4 w_2
\rho_4^\prime +  w_4^\prime w_2
\rho_4 + w_4 w_2^\prime \rho_4 + {1\over 2} w_2^2  \rho_4^{\prime\prime\prime}
 \cr & + {3\over 2} w_2 w_2^\prime  \rho_4^{\prime\prime} + {1\over 2}
{w_2^\prime}^2  \rho_4^{\prime} + {11\over 10} w_2 w_2^{\prime\prime}
\rho_4^{\prime} \cr & + {3\over 10} w_2^\prime w_2^{\prime\prime} \rho_4 +
{3\over 10} w_2 w_2^{\prime\prime\prime} \rho_4 +  {1\over 20}
\rho_4^{\prime\prime\prime\prime\prime\prime\prime},
\cr
\delta(\rho_4)\,h_2  =&  {9\over 10} h_2^{\prime\prime
\prime} \rho_4 -{3\over 100} h_4
\rho_4^{\prime\prime \prime\prime\prime}  +{1\over 20} h_4^\prime
\rho_4^{\prime
\prime\prime\prime} -{3\over 50} h_4^{\prime\prime}
\rho_4^{\prime\prime\prime} +{3\over 50} h_4^{\prime\prime\prime}
\rho_4^{\prime\prime} - {1\over 20} h_4^{\prime\prime\prime\prime}
\rho_4^{\prime} \cr & +{3\over 100} h_4^{\prime\prime\prime\prime\prime}
\rho_4
-{3\over 10} w_2 h_4 \rho_4^{\prime\prime\prime}
-{3\over 5} w_2^\prime h_4 \rho_4^{\prime\prime}
+{1\over 5} w_2 h_4^\prime \rho_4^{\prime\prime}
-{1\over 5} w_2 h_4^{\prime\prime} \rho_4^{\prime}
-{3\over 5} w_2^{\prime\prime} h_4 \rho_4^{\prime}
\cr & +{3\over 5} w_2^\prime h_4^{\prime\prime} \rho_4 +{3\over 10} w_2
h_4^{\prime\prime\prime} \rho_4
 + w_4 h_4 \rho_4^{\prime} + w_4 h_4^{\prime} \rho_4 + {1\over 2} h_3
w_3^{\prime}\rho_4 + {1\over 2} h_3 w_3 \rho_4^{\prime},\cr
\delta(\rho_4)\,h_3  =&  {1\over 10} (h_3 \rho_4^{\prime\prime
\prime}  -3 h_3^\prime \rho_4^{\prime
\prime} +5  h_3^{\prime\prime}
\rho_4^{\prime} -5 h_3^{\prime\prime\prime}
\rho_4)  \cr & - {1\over 2} w_2 h_3^{\prime} \rho_4 - {1\over 2} w_2^{\prime}
h_3
\rho_4 - {3\over 4} w_3 h_4^\prime \rho_4 + {3\over 4} w_3 h_4
\rho_4^\prime,\cr
\delta(\rho_4)\,h_4  =&
3 h_2^\prime \rho_4 - h_2 \rho_4^\prime + {\dot \rho_4} \cr &
-{1\over 10}( h_4 \rho_4^{\prime\prime\prime}  - 2 h_4^\prime
\rho_4^{\prime\prime} -2 h_4^{\prime\prime} \rho_4^{\prime} +
 h_4^{\prime\prime\prime} \rho_4  +
h_4^\prime w_2 \rho_4 - h_4 w_2 \rho_4^\prime).\cr} $$

\vskip 0.5cm
\noindent
{\bf Appendix B}:

\noindent
Looking at the gauge symmetries
preserving the gauge as defined in Eq.(44) for $sl(10,R)$, we get the
following Virasoro transformations of the $w_i$'s, $2\le i \le 10$:
$$
\eqalign{
\delta(\rho_2) w_2 &= 2 w_2 \rho_2^\prime +  w_2^\prime \rho_2 + {165\over
2}\rho_2^{\prime\prime\prime},\cr
\delta(\rho_2) w_3 &= 3 w_3 \rho_2^\prime +  w_3^\prime \rho_2,\cr
\delta(\rho_2) w_4 &= 4 w_4 \rho_2^\prime +  w_4^\prime \rho_2 + {266\over 5}
w_2
\rho_2^{\prime\prime\prime},\cr
}$$
\vfill\eject
$$\eqalign{
\delta(\rho_2) w_5 &= 5 w_5 \rho_2^\prime +  w_5^\prime \rho_2 + {83\over 2}
w_3
\rho_2^{\prime\prime\prime},\cr
\delta(\rho_2) w_6 &= 6 w_6 \rho_2^\prime +  w_6^\prime \rho_2 + {185\over 6}
w_4
\rho_2^{\prime\prime\prime}\cr
  &\quad +{52\over 3}  w_2 \rho_2^{\prime\prime\prime\prime\prime}
 -{130\over 3}  w_2^{\prime} \rho_2^{\prime\prime\prime\prime}
 +{130\over 3}  w_2^{\prime\prime} \rho_2^{\prime\prime\prime},\cr
\delta(\rho_2) w_7 &= 7 w_7 \rho^\prime +  w_7^\prime \rho_2 + {235\over 11}
w_5
\rho_2^{\prime\prime\prime}\cr
  &\quad +{119\over 11}  w_3 \rho_2^{\prime\prime\prime\prime\prime}
 -{595\over 33}  w_3^{\prime} \rho_2^{\prime\prime\prime\prime}
 +{170\over 11}  w_3^{\prime\prime} \rho_2^{\prime\prime\prime},\cr
\delta(\rho_2) w_8 &= 8 w_8 \rho_2^\prime +  w_8^\prime \rho_2 + {173\over 13}
w_6
\rho_2^{\prime\prime\prime}\cr
  &\quad +{762\over 143}  w_4 \rho_2^{\prime\prime\prime\prime\prime}
 -{1905\over 286}  w_4^{\prime} \rho_2^{\prime\prime\prime\prime}
 +{635\over 429}  w_4^{\prime\prime} \rho_2^{\prime\prime\prime} \cr
  &\quad +{23\over 33}  w_2 \rho_2^{\prime\prime\prime\prime\prime\prime\prime}
 -{161\over 33}  w_2^{\prime} \rho_2^{\prime\prime\prime\prime\prime\prime}
 +{483\over 55}  w_2^{\prime\prime} \rho_2^{\prime\prime\prime\prime\prime}
 -{161\over 33}  w_2^{\prime\prime\prime} \rho_2^{\prime\prime\prime\prime}
 +{23\over 33}  w_2^{\prime\prime\prime\prime} \rho_2^{\prime\prime\prime}
,\cr
\delta(\rho_2) w_9 &= 9 w_9 \rho_2^\prime +  w_9^\prime \rho_2 + {173\over 13}
w_7
\rho_2^{\prime\prime\prime}\cr
  &\quad +{762\over 143}  w_5 \rho_2^{\prime\prime\prime\prime\prime}
 -{1905\over 286}  w_5^{\prime} \rho_2^{\prime\prime\prime\prime}
 +{635\over 429}  w_5^{\prime\prime} \rho_2^{\prime\prime\prime} \cr
  &\quad +{23\over 33}  w_3 \rho_2^{\prime\prime\prime\prime\prime\prime\prime}
 -{161\over 33}  w_3^{\prime} \rho_2^{\prime\prime\prime\prime\prime\prime}
 +{483\over 55}  w_3^{\prime\prime} \rho_2^{\prime\prime\prime\prime\prime}
 -{161\over 33}  w_3^{\prime\prime\prime} \rho^{\prime\prime\prime\prime}
 +{23\over 33}  w_3^{\prime\prime\prime\prime} \rho_2^{\prime\prime\prime},\cr
\delta(\rho_2) w_{10} &= 10 w_{10} \rho_2^\prime +  w_{10}^\prime \rho_2 +
{173\over
13} w_8 \rho_2^{\prime\prime\prime}\cr
  &\quad +{762\over 143}  w_6 \rho_2^{\prime\prime\prime\prime\prime}
 -{1905\over 286}  w_6^{\prime} \rho_2^{\prime\prime\prime\prime}
 +{635\over 429}  w_6^{\prime\prime} \rho_2^{\prime\prime\prime} \cr
  &\quad +{23\over 33}  w_4 \rho_2^{\prime\prime\prime\prime\prime\prime\prime}
 -{161\over 33}  w_4^{\prime} \rho_2^{\prime\prime\prime\prime\prime\prime}
 +{483\over 55}  w_4^{\prime\prime} \rho_2^{\prime\prime\prime\prime\prime}
 -{161\over 33}  w_4^{\prime\prime\prime} \rho_2^{\prime\prime\prime\prime}
 +{23\over 33}  w_4^{\prime\prime\prime\prime} \rho_2^{\prime\prime\prime}\cr
  &\quad +{23\over 33}  w_2
\rho_2^{\prime\prime\prime\prime\prime\prime\prime\prime\prime}
 -{161\over 33}  w_2^{\prime}
\rho_2^{\prime\prime\prime\prime\prime\prime\prime\prime}
 +{483\over 55}  w_2^{\prime\prime}
\rho_2^{\prime\prime\prime\prime\prime\prime\prime}
 -{161\over 33}  w_2^{\prime\prime\prime}
\rho_2^{\prime\prime\prime\prime\prime\prime}
 +{23\over 33}  w_2^{\prime\prime\prime\prime}
\rho_2^{\prime\prime\prime\prime\prime}
 \cr &\quad-{161\over 33}  w_2^{\prime\prime\prime\prime}
\rho_2^{\prime\prime\prime\prime\prime}
 +{23\over 33}  w_2^{\prime\prime\prime\prime\prime\prime}
\rho_2^{\prime\prime\prime}
.\cr
}
$$

\vskip 0.7cm
\noindent
{\bf Appendix C}:

\noindent As mentioned in the text, the Virasoro variation of $h_i$ with $i>2$
is that of a current with dimension $i$,
$$ \delta(\rho_2) h_{i,i>2} = -(i-1)h_i\,\rho_2^\prime +
h_i^\prime \rho_2. $$
However, the variation of  $h_2$ involves higher derivative terms
for $N \ge 4$. Denoting
$${\cal L}_{\rho_2}\,h_2 = - h_2\, \rho_2^\prime\, +
\rho_2\,h_2^\prime + {\dot \rho_2},$$ we obtain for $N\ge 4$
$$\delta({\rho_2})\,h_2  = {\cal L}_{\rho_2}\,h_2 + \Gamma_N,$$
where $\Gamma_N$ is vanishing in the finite subgroup of diffeomorphism, in
which case $\rho_2^{\prime\prime\prime} = 0$. In addition to
$\delta({\rho_2})\,h_2$ given in Eq.(59), some explicit expressions are given
below $$ \eqalign{
&\Gamma_5  = - {16\over 5} h_4\, \rho_2^{\prime\prime\prime},\cr
&\Gamma_6  = - {37\over 5} h_4\, \rho_2^{\prime\prime\prime}-{5 \over 7} h_6\,
\rho_2^{\prime\prime\prime\prime\prime}-{5 \over 7} h_6^\prime\,
\rho_2^{\prime\prime\prime\prime}-{10 \over 63} h_6^{\prime\prime}\,
\rho_2^{\prime\prime\prime},\cr
&\Gamma_7  = - {14} h_4\,
\rho_2^{\prime\prime\prime} -{57 \over 14} h_6\,
\rho_2^{\prime\prime\prime\prime\prime} -{57 \over 14} h_6^\prime\,
\rho_2^{\prime\prime\prime\prime}-{19 \over 21} h_6^{\prime\prime}\,
\rho_2^{\prime\prime\prime},\cr
&\Gamma_8  =  - {47\over 2} h_4\,
\rho_2^{\prime\prime\prime} -{96 \over 7} h_6\,
\rho_2^{\prime\prime\prime\prime\prime} -{96 \over 7} h_6^\prime\,
\rho_2^{\prime\prime\prime\prime}-{64 \over 21} h_6^{\prime\prime}\,
\rho_2^{\prime\prime\prime}\cr & \qquad -{7 \over 12} h_8\,
\rho_2^{\prime\prime\prime\prime\prime\prime\prime} -{7 \over 6} h_8^\prime\,
\rho_2^{\prime\prime\prime\prime\prime\prime}-{21 \over 26}
h_8^{\prime\prime}\,
\rho_2^{\prime\prime\prime\prime\prime}-{35 \over 156}
h_8^{\prime\prime\prime}\, \rho_2^{\prime\prime\prime\prime}-{35 \over 1716}
h_8^{\prime\prime\prime\prime}\, \rho_2^{\prime\prime\prime},\cr
&\Gamma_9  = - {182\over 5} h_4\,
\rho_2^{\prime\prime\prime} -{71 \over 2} h_6\,
\rho_2^{\prime\prime\prime\prime\prime} -{71 \over 2} h_6^\prime\,
\rho_2^{\prime\prime\prime\prime}-{71 \over 9} h_6^{\prime\prime}\,
\rho_2^{\prime\prime\prime}\cr & \qquad -{68 \over 15} h_8\,
\rho_2^{\prime\prime\prime\prime\prime\prime\prime} -{136 \over 15}
h_8^\prime\,
\rho_2^{\prime\prime\prime\prime\prime\prime}-{408 \over 65}
h_8^{\prime\prime}\,
\rho_2^{\prime\prime\prime\prime\prime}-{68 \over 39}
h_8^{\prime\prime\prime}\, \rho_2^{\prime\prime\prime\prime}-{68 \over 429}
h_8^{\prime\prime\prime\prime}\, \rho_2^{\prime\prime\prime},\cr
&\Gamma_{10}  = - {256\over 5} h_4\,
\rho_2^{\prime\prime\prime} -{78} h_6\,
\rho_2^{\prime\prime\prime\prime\prime} -{78} h_6^\prime\,
\rho_2^{\prime\prime\prime\prime}-{52 \over 3} h_6^{\prime\prime}\,
\rho_2^{\prime\prime\prime}\cr & \qquad -{299 \over 15} h_8\,
\rho_2^{\prime\prime\prime\prime\prime\prime\prime} -{598 \over 15}
h_8^\prime\,
\rho_2^{\prime\prime\prime\prime\prime\prime}-{138 \over 5}
h_8^{\prime\prime}\,
\rho_2^{\prime\prime\prime\prime\prime}-{23 \over 3}
h_8^{\prime\prime\prime}\, \rho_2^{\prime\prime\prime\prime}-{23 \over 33}
h_8^{\prime\prime\prime\prime}\, \rho_2^{\prime\prime\prime}\cr  &
\qquad -{27 \over 55} h_{10}\,
\rho_2^{\prime\prime\prime\prime\prime\prime\prime\prime\prime} -{81 \over
35} h_{10}^\prime\,
\rho_2^{\prime\prime\prime\prime\prime\prime\prime\prime}-{324 \over 187}
h_{10}^{\prime\prime}\,
\rho_2^{\prime\prime\prime\prime\prime\prime\prime}-{189 \over 187}
h_{10}^{\prime\prime\prime}\,
\rho_2^{\prime\prime\prime\prime\prime\prime}\cr  &
\qquad -{567 \over 1870}
h_{10}^{\prime\prime\prime\prime}\,
\rho_2^{\prime\prime\prime\prime\prime}-{81 \over 1870}
h_{10}^{\prime\prime\prime\prime\prime}\,
\rho_2^{\prime\prime\prime\prime}-{27 \over 12155}
h_{10}^{\prime\prime\prime\prime\prime\prime}\, \rho_2^{\prime\prime\prime}.
}$$
We notice that $\Gamma_N$ involves only even dimensional $h_i$'s, e.g. $h_4,
h_6, \dots$. As already mentioned in the $sl(4,R)$ case, these higher
derivative
terms can be removed by a redefinition of $H_2$ that introduces coupling terms.
For example,
$$\eqalign{
N=5:\,&H_2 = h_2 - {3\over 2}h_3^\prime +{3\over 5} \big( \partial^2 - {8\over
15}w_2\big)h_4 - {1\over 14} h_5^{\prime\prime\prime}, \cr
N=6:\,&H_2 = h_2 - 2 h_3^\prime +{6\over 5} \big( \partial^2 - {37\over
105}w_2\big)h_4 - {2\over 7}h_5^{\prime\prime\prime} \cr
&\qquad +{1\over 42} \big(
\partial^4 - {12\over 7}({2\over 9}w_2\partial^2 + {w_2}^\prime\partial +
w_2^{\prime\prime}-4{w_2}^2) \big) h_6, \cr
N=7:\,&H_2 = h_2 - {5 \over 2} h_3^\prime + 2 \big( \partial^2 - {1\over
4}w_2\big)h_4 - {5\over 7}h_5^{\prime\prime\prime} \cr
&\qquad +{5\over 42} \big(
\partial^4 + {171\over 140}({2\over 9}w_2\partial^2 +  {w_2}^\prime\partial +
w_2^{\prime\prime}-4{w_2}^2) \big) h_6 - {1\over
132}h_7^{\prime\prime\prime\prime\prime},\cr }$$
will lead to $\delta({\rho_2})\,h_2  = {\cal L}_{\rho_2}\,h_2$. A
general all order expression for $H_2$ is under investigation.

\vfill\eject
\bigskip
\centerline {References}
\bigskip
\item{[1]} A.M. Polyakov, Mod. Phys.
Lett.{\bf A2} (1987) 893; See also {\sl Two-Dimensional quantum gravity and
Superconductivity at high $T_c$} in $\underline{\hbox{Fields, Strings and
Critical Phenomena}}$, eds. E. Br\'ezin and J. Zinn-Justin, Les Houches 1988,
Elsevier Science Publishers, Amsterdam, (1989).
\smallskip
\item{[2]}
A.H. Chamseddine and M. Reuter, Nucl. Phys. {\bf B317} (1989) 757;
\hfill\break
A. Alekseev and S. Shatashvili, Nucl. Phys. {\bf B323} (1989) 719;
\hfill\break
M. Bershadsky and H. Ooguri, Comm. Math. Phys. {\bf 126} (1989) 49;
\hfill\break
A.M. Polyakov, Mod. Phys. Lett.{\bf A6} (1991) 635;
\hfill\break
K. Isler and J-M. Lina, Nucl. Phys. {\bf B358} (1991) 713;
\hfill\break
D. Kutasov, E. Martinec and N. Seiberg,  preprint PUPT- 1923, RU-91-49 and
references therein.
\hfill\break
For the approach in the conformal gauge, see E. D'Hoker, {\sl Lecture notes on
2-d quantum gravity and Liouville theory}, preprint UCLA/91/TEP/35 (1991), and
references therein.
\smallskip \item{[3]} M.R. Douglas, Phys. Lett. B238 (1990)
176; \hfill\break
M.A. Awada and Z. Qiu, Phys. Lett. B245 (1990) 85;
\hfill\break
M.A. Awada and S.J. Sin, Institute of fundamental theory preprint UFIFT
HEP-91-3 (1991).
\smallskip
\item{[4]} V.G. Knizhnik, A.M. Polyakov and A.B.
Zamolodchikov, Mod. Phys. Lett. {\bf A3} (1988) 819;
\hfill\break
J. Distler and H. Kawai,  Nucl. Phys. {\bf B321} (1989) 509.
\smallskip
\item{[5]} A.M. Polyakov,Int. J. Mod. Phys. {\bf A5} (1990) 833.
\smallskip
\item{[6]} A. Bilal, V.V.  Fock and I.I. Kogan, Nucl. Phys. {\bf B359} (1991)
635, and references therein;
\hfill\break
H. Ooguri, K. Schoutens, A. Sevrin and P. Van Nieuwenhuizen, Stony Brook
preprint ITP-SB-91-16, RIM-764.
\hfill\break
Other approaches to $w_N$ gravity can be found in the following papers
\hfill\break
C.M. Hull, Phys. Lett. B240 (1990) 110; {\sl ibid.} B269 (1991)
257;
\hfill\break
K. Schoutens, A. Sevrin and P. Van Nieuwenhuizen, Stony Brook
preprint ITP-SB-91-21; Phys. Lett. B243 (1990) 295.
\hfill\break
R. Parthasarathy and K.S. Viswanathan, Simon Fraser Univ. preprint (1991).
\smallskip
\item{[7]} A. Das, W-J. Huang and S. Roy, Rochester preprint
UR1197, ER13065-652. \smallskip
\item{[8]} A.M. Polyakov and P. Wiegmann, Phys. Lett. B131 (1983) 121; Phys.
Lett. B141 (1984) 223.
\smallskip
\item{[9]} J.M. Lina and P.K. Panigrahi, Mod. Phys. Lett. {\bf A6} (1991) 3517.
\smallskip
\item{[10]}
A. Gerasimov, A. Levin and A. Marshakov, Nucl. Phys. {\bf B317} (1989) 757.
\smallskip
\item{[11]} A.B. Zamolodchikov, Teor. Mat. Fiz 65 (1985) 1205;
\hfill\break
V.A. Fateev and S.L. Lukyanov, Int. Jour. Mod. Phys. {\bf A3} (1988) 507.
\smallskip
\item{[12]} A. Bilal, {\sl Introduction to $W$-algebras} CERN-TH. preprint
6083/91,
LPTENS 91/10 (1991).
\smallskip
\item{[13]} J. Balog, L. Feh\'er, L. O'Raifeartaigh and A.
Wipf, Phys. Lett. B227 (1989) 214;
\hfill\break
L. Feh\'er, L. O'Raifeartaigh, P. Ruelle, I.
Tsusui and A. Wipf,  preprint U.de M.-LPN-TH.71, DIAS-STP-91-29 (1991) and
references therein. \smallskip
\item{[14]} F. Bais, P. Bouwknegt, M. Surridge and K. Schoutens, Nucl. Phys.
{\bf
B304} (1988) 348, {\sl ibid.} 371.
\smallskip
\item{[15]} A. Bilal and J.L. Gervais, Nucl. Phys. {\bf B314} (1989) 646, {\sl
ibid.} {\bf
B318} (1989) 579; \smallskip
\item{[16]} H. Verlinde,  Nucl. Phys. {\bf B337} (1990) 652;
\hfill\break
A. Bilal, Phys. Lett. B267 (1991) 487.
\smallskip
\item{[17]} K. Yamagishi,  Phys. Lett. B205 (1988) 466;
\hfill\break
P. Mathieu,  Phys. Lett. B208 (1988) 101;
\hfill\break
Q. Wang, P.K. Panigrahi, U. Sukhatme and W-Y. Keung,  Nucl. Phys. {\bf B344}
(1990)
196;
\hfill\break
I. Bakas, Phys. Lett. B228 (1989) 57.
\smallskip
\item{[18]} J.L. Gervais,  Phys. Lett. B160 (1985) 277;
\hfill\break
F. Magri, Jour. of Math. Phys. 19 (1978) 1156.
\hfill\break See also P.J. Olver, {\sl Applications of Lie groups to
differential equations}, Springer, Berlin (1986);
\hfill\break
A. Das, {\sl Integrable Models}, World Scientific (1989).
\smallskip
\item{[19]} W. Scherer, Lett. Math. Phys. {\bf 17} (1989) 45.
\smallskip
\item{[20]} A. Das and S. Roy, Rochester preprint UR-1151 (1990).
\smallskip
\item{[21]} B.L. Feigin and D.B. Fuchs, Func. Anal. \& Appl. {\bf 17} (1982)
114;
\hfill\break
P. Di Francesco, C. Itzykson and J-B. Zuber, Princeton preprint SPhT/90-149,
PUPT-1211.
\smallskip
\item{[22]} A.B. Zamolodchikov,  preprint ITEP 89-84 (1989).
\smallskip
\item{[23]} Y. Matsuo,  Nucl. Phys. {\bf B342} (1990) 643.
\smallskip
\item{[24]} V. Drinfeld and V. Sokolov, Jour. Sov. Math. 30 (1985) 1975;
\hfill\break
I.M. Gelfand and L.A. Dikki, Funct. Anal. Appl.{\bf 10} (1976) 259;
\hfill\break
A.M. Adler, Invent. Math. {\bf 50} (1979) 219.
\smallskip
\item{[25]} A.M. Adler, see Ref.[24].
\smallskip
\item{[26]} K. Isler and C.A. Trugenberger, Phys. Rev. Lett. 63 (1989) 834;
\hfill\break
E. D'Hoker, preprint UCLA/91/TEP/8.
\smallskip
\item{[27]}  This analysis has been made possible by the use of symbolic
programing with {\sl Mathematica}. Formula (53) has been checked up to $N=40$,
using a program available by e-mail from the following adress:
$lina@lps.umontreal.ca$.
\smallskip
\item{[28]} J. Pawelczyk, preprint WIS-90/28/june (1990).

\end